\documentclass[superscriptaddress,twocolumn,amsmath,amssymb, aps, prl]{revtex4-2}

\usepackage{graphicx}
\usepackage{dcolumn}
\usepackage{bm}
\usepackage[colorlinks=true]{hyperref}
\usepackage[version=3]{mhchem}
\usepackage{xcolor}
\usepackage{tikz}
\usepackage{braket}
\usepackage{gensymb}

\definecolor{ricky}{cmyk}{0, 0.7808, 0.4429, 0.1412}
\definecolor{dftblue}{RGB}{50,100,170}
\definecolor{zhanggreen}{RGB}{35,130,80}
\definecolor{hirakiorange}{RGB}{205,115,35}
\definecolor{morawetzred}{RGB}{150,45,45}
\definecolor{axisgray}{RGB}{90,90,90}
\definecolor{basegray}{RGB}{180,180,180}

\newcommand{\thor}{$^{229}$Th}

\begin{document}

\title{Defect assignment of the clock site in \(^{229}\)Th:CaF\(_2\)}

\author{Daniel A. Rehn}
\affiliation{Computational Physics Division, Los Alamos National Laboratory, Los Alamos, NM 87545, USA}

\author{Harry W. T. Morgan}
\affiliation{Department of Chemistry, University of Manchester, Manchester, UK}

\author{Harris E. Mason}
\affiliation{Chemistry Division, Los Alamos National Laboratory, Los Alamos, NM 87545, USA}

\author{H. B. Tran Tan}
\affiliation{Computational Physics Division, Los Alamos National Laboratory, Los Alamos, NM 87545, USA}

\author{Ricky Elwell}
\affiliation{Department of Physics and Astronomy, University of California, Los Angeles, Los Angeles, CA, USA}

\author{Igor M. Savukov}
\affiliation{Materials Physics and Applications Division, Los Alamos National Laboratory, Los Alamos, NM 87545, USA}

\author{Michael J. Martin}
\affiliation{Materials Physics and Applications Division, Los Alamos National Laboratory, Los Alamos, NM 87545, USA}

\author{Andrei Derevianko}
\affiliation{Department of Physics, University of Nevada, Reno, Reno, NV, USA}

\author{Eric R. Hudson}
\affiliation{Department of Physics and Astronomy, University of California, Los Angeles, Los Angeles, CA, USA}
\affiliation{Challenge Institute for Quantum Computation, University of California, Los Angeles, Los Angeles, CA, USA}
\affiliation{Center for Quantum Science and Engineering, University of California, Los Angeles, Los Angeles, CA, USA}

\begin{abstract}
The performance of solid-state \(^{229}\)Th nuclear clocks depends sensitively
on the microscopic environment of the thorium nucleus in the host crystal. Here
we reassess the dominant quadrupole-split thorium site in
\(^{229}\)Th:CaF\(_2\), which has been assigned to a thorium dimer in recent
spectroscopic work. Thermodynamic estimates, density functional theory
calculations, and electric-field-gradient comparisons instead favor an isolated
\ce{Th^4+} substitution on a \ce{Ca^2+} site charge-compensated by two nearby
fluorine interstitials in a relaxed \(90^\circ\) motif. The same calculation
identifies a higher-energy mixed-shell interstitial motif as a plausible minor
site. The clock-active quadrupole-split site is therefore controlled by local
fluoride compensation rather than unavoidable thorium aggregation. This defect
assignment also has implications for achievable linewidths and provides a
microscopic basis for reducing broadening in solid-state nuclear clocks.
\end{abstract}

\maketitle
The low-energy isomeric transition in \thor{} offers a paradigm shift in
timekeeping~\cite{PieTam03} by enabling a solid-state nuclear
clock~\cite{Rellergert2010}, with prospective sensitivity to physics beyond the
Standard Model~\cite{DarkMatterSignal}. Recent breakthroughs in laser excitation
of \thor{} in wide-band-gap
crystals~\cite{Tiedau2024-caf2,Elwell2024-lisaf,Zhang2024-ThF4} have enabled
high-resolution spectroscopy of the nuclear
transition~\cite{Zhang2024-Th229Comb,Hiraki2025}, studies of potential clock
systematics~\cite{Higgins2025-Temperature-dependence-ThCaF2,Terhune2025-photo-induced,Ooi2026},
extension to low-band-gap platforms~\cite{Elwell2025-ThO2}, and early
demonstrations of nuclear clocks~\cite{ToscaniDeCol2026clock,Huang2026clock}.

The performance of a solid-state nuclear clock depends critically on the local
environment of the \thor{} nucleus in the host material.
Among candidate materials, Th-doped \ce{CaF2} has been studied most extensively.
It combines high VUV transparency with \thor{} incorporation at densities
sufficient for resolved nuclear
spectroscopy.~\cite{Tiedau2024-caf2,Zhang2024-Th229Comb,Hiraki2025}. In
\ce{CaF2}, \thor{}$^{4+}$ substitutes for \ce{Ca^2+}. The
charge difference between the cations must be balanced by other charged defects,
and the resulting defect complex determines the spectroscopic properties of
\thor{} and the ultimate stability and accuracy attainable in a clock.
Therefore, identifying the dominant thorium defect configuration is critical for
understanding and developing \thor{}:\ce{CaF2} nuclear clocks.

Recent laser M\"ossbauer spectroscopy of \thor{}:\ce{CaF2} resolved multiple
thorium environments. The two strongest spectral components were assigned to a
zero-electric field gradient (EFG) ``O center'' and to a thorium-dimer ``D
center'' without nearby charge compensation~\cite{Hiraki2025}. Because the O
center exhibits a broad linewidth, subsequent clock operation has focused on the
strongest quadrupole-split
site~\cite{Morawetz2026,ToscaniDeCol2026clock,Huang2026clock}. The composition
and structure of that site are therefore crucial for clock development: if the
clock transition is primarily hosted by Th--Th aggregates, then linewidths and
frequency shifts are tied directly to concentration-dependent many-dopant
complexes. If the spectrum comes instead from a locally charge-compensated
single-Th site, then material optimization should focus on controlling nearby
charge-compensation defects, strain, and interconversion between defect
complexes.

Here we show that the clock site is unlikely to be due to
Th--Th dimers in the dilute clock crystals. The thermodynamics favor local
charge compensation, and the observed quadrupole spectrum is most consistent
with an isolated \ce{Th^4+} on a \ce{Ca^2+} site charge-compensated by two
nearby \ce{F-} interstitials. This assignment is consistent with prior
formation-energy calculations favoring local fluoride
compensation~\cite{Dessovic2014} and with recent EXAFS measurements that
identify a locally fluorinated thorium environment~\cite{Takatori2025}. The
assignment rests on three tests: dimer and local-compensation thermodynamics,
the symmetry-reduced set of two-F$_i$ motifs in density-functional-theory
calculations, and the resulting quadrupole branch patterns.

\begin{figure*}[t!]\centering
\includegraphics[width=\linewidth]{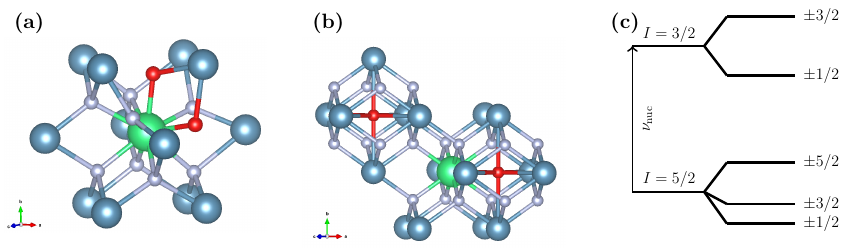}
\caption{Representative single-Th charge-compensation motifs in CaF$_2$. In
panels (a) and (b), Th is green, Ca is blue, lattice F atoms are white, and
added interstitial F atoms are red. (a) The thermodynamically preferred
90$^\circ$ two-F$_i$ motif. (b) The mixed first/second-shell 1st-2nd obtuse
motif, which is the next most likely two-F$_i$ defect in
Table~\ref{tab:pbe_interstitial_prl}. (c) Nuclear energy splitting produced by a
nonzero EFG. \label{fig:motifs}}
\end{figure*}

The proposed D center consists of two \ce{Th^4+} ions on adjacent
\ce{Ca^2+} sites without local charge compensation.
It therefore has a net +4 charge and no nearby compensating defect.
It was assigned as the principal defect because the calculated dimer EFG
\((V_{ZZ}=95~\text{V \AA}^{-2}, \eta=0.6)\) was close to the measured
values \((V_{ZZ}=106.3(8)~\text{V \AA}^{-2}, \eta=0.60(1))\)~\cite{Hiraki2025}, where numbers in
parentheses denote 68\% confidence intervals.
Earlier calculations had predicted the lowest-energy defect to be a single
\ce{Th^4+} on a \ce{Ca^2+} site with two \ce{F-} interstitials,
Th$_\mathrm{Ca}^{\bullet\bullet}+2\mathrm{F}_i^\prime$ in Kr\"oger--Vink
notation, but the predicted EFG
\((V_{ZZ}=223.0~\text{V \AA}^{-2})\) differed substantially from experiment,
so that structure was discounted~\cite{Dessovic2014}.

Thermodynamics point in the opposite direction. The calculated energy to form
the proposed D center from two separated thorium defects is
\(+0.56\,\mathrm{eV}\), consistent with repulsion between
nearby \ce{Th^4+} substitutions. Configurational entropy also disfavors
dimerization in a dilute crystal: randomly distributed thorium ions very rarely
occupy nearest-neighbor Ca sites. For the \ce{^{229}Th} concentration of the X2
clock crystal~\cite{Hiraki2025}, \(N_{\rm Ca}\simeq10^4\) Ca sites are available
for two Th dopants. The ideal-mixing estimate in Supplemental
Sec.~\textcolor{red}{XI} gives an additional dimer-formation free-energy cost of
\(0.43\,\mathrm{eV}\) at 300~K and
\(2.41\,\mathrm{eV}\) at 1691~K, near the
melting temperature of \ce{CaF2}. Thus the relevant dimer association free
energy is positive even before considering the distant defect required for
charge neutrality. One remaining piece of evidence for the dimer assignment is
the electron-microscopy observation of Th clustering in Ref.~\cite{Hiraki2025}.
This observation, however, was made in a much more highly doped crystal with a thorium density of
\(2.6\times10^{20}\) cm\(^{-3}\), or 1.1 at.\%, much higher than the
\ce{^{229}Th}:\ce{CaF2} clock crystals \((5\times10^{18}\) cm\(^{-3}\), or 0.02
at.\%)~\cite{Zhang2024}.  Rare-earth-doped \ce{CaF2} illustrates why this
concentration difference matters: in Er:\ce{CaF2}, low-concentration dopants are
isolated, whereas high-concentration dopants form rare-earth clusters bound to
charge-compensating Ca vacancies~\cite{DAcapito2016ErCaF2,Wang25ErCaF2}. Thus
clustering observed in a highly-doped microscopy sample supports the possibility
of Th association in \ce{CaF2}, but it does not establish that bare Th--Th
dimers dominate the dilute clock crystals.

\begin{table*}[t]
\refstepcounter{table}
\label{tab:pbe_interstitial_prl}
\noindent\begin{minipage}{\textwidth}
\raggedright
\textbf{Table \thetable.}
PBE/SOC comparison of the same-stoichiometry two-F$_i$ defect family using a
\(5\times5\times5\) supercell of the primitive 3-atom \ce{CaF2} unit cell. All
rows have composition \({\rm Th}_1{\rm Ca}_{124}{\rm F}_{252}\), so the energy
differences can be compared directly within this family, but should not be
compared directly with the bare-dimer association energy discussed above.
The principal EFG component \(V_{ZZ}\) is in V/\AA$^2$, \(\Delta E_i\) is
referenced to the \(90^\circ\) F$_i$ pair, and \(p_i\) is
the degeneracy-weighted restricted-equilibrium fraction at 1400 K, the reported
CaF$_2$ superionic-transition scale.~\cite{CazorlaErrandonea2014CaF2Superionic}.
These fractions compare local motifs within this two-F$_i$ family and are not
absolute growth-yield predictions.
\end{minipage}

\vspace{0.25em}
\centering
\scriptsize
\begin{ruledtabular}
\begin{tabular}{lrrrrr}
Defect model & \(V_{ZZ}\) & \(\eta\)
& \(g_i\) & \(\Delta E_i\) (eV) & \(p_i(1400~{\rm K})\)  \\
\hline
90 deg & 107.650 & 0.173 & 12 & 0.00 & 0.9970  \\
1st-2nd obtuse & -309.049 & 0.156 & 24 & 0.86 & \(1.56\times10^{-3}\)  \\
180 deg & -348.163 & 0.000 & 3 & 0.69 & \(8.39\times10^{-4}\)  \\
1st-2nd acute & -223.028 & 0.493 & 24 & 0.98 & \(5.77\times10^{-4}\)  \\
2nd-2nd tetra & -56.157 & 0.963 & 12 & 1.62 & \(1.46\times10^{-6}\)  \\
2nd-2nd edge & -37.264 & 0.793 & 12 & 1.73 & \(5.76\times10^{-7}\)  \\
2nd-2nd opp. & -117.081 & 0.000 & 4 & 1.61 & \(5.45\times10^{-7}\) \\
\end{tabular}
\end{ruledtabular}
\end{table*}

Having established that bare dimers are thermodynamically disfavored at the
dilute thorium concentrations of the clock crystals, the next question is which
locally compensated single-Th structure is expected once fluoride compensation
is present. This narrower comparison is shown in
Table~\ref{tab:pbe_interstitial_prl}, which ranks neutral, same-stoichiometry
two-F$_i$ motifs. Calculations here and in previous work~\cite{Dessovic2014}
identify Th$_\mathrm{Ca}^{\bullet\bullet}+2\mathrm{F}_i^\prime$ as the
lowest-energy local compensation family, with the two nearby fluoride
interstitials initially placed in a 90\degree{} motif. Forming this cluster from
isolated point defects is exothermic by \(6.15\,\mathrm{eV}\),
reflecting the electrostatic stabilization of local charge compensation. A
conservative ideal-mixing estimate gives an entropy penalty below this
association enthalpy. At the superionic transition of 1400
K~\cite{CazorlaErrandonea2014CaF2Superionic} this upper-bound entropy penalty is
about \(2.98\,\mathrm{eV}\) after including the 90\degree{}
orientational degeneracy \(g_i=12\); even at the 1691 K melt temperature it
remains about \(3.60\,\mathrm{eV}\), still well below the
\(6.15\,\mathrm{eV}\) association enthalpy. Details of the
association-energy and entropy estimates are given in Supplemental
Sec.~\textcolor{red}{XI}. The relevant energy scale for deciding between local
charge compensation and a bare dimer is therefore the large association energy
for bringing compensating F$_i$ defects to Th, not the sub-eV differences
between already compensated 2F$_i^\prime$ geometries in
Table~\ref{tab:pbe_interstitial_prl}. Thus, the expected equilibrium population
is not dominated by bare Th substitutions or dimers, but locally compensated Th
centers.

This local-compensation argument does not uniquely assign the zero-EFG O center.
Because the pristine \ce{CaF2} Ca site is cubic and has zero EFG, a near-zero
line could arise from distant compensation, dynamic averaging among equivalent
F$_i$ orientations, radiation-induced dissociation of local compensators, or
larger complexes outside the static local models considered here; recent
embedded-cluster calculations likewise emphasize that vacancy-related
compensation can substantially alter the local electronic structure of
\thor{}:\ce{CaF2}~\cite{Nalikowski2024Embedding}.
Rare-earth-doped \ce{CaF2} provides a precedent: ultraviolet radiation can
dissociate M$_\mathrm{Ca}^{\bullet}$ centers from their local F$_i^\prime$
charge compensators, creating O-center-like sites~\cite{Twidell1963}.
Temperature- and irradiation-dependent changes in the O-center fraction would
therefore be a useful experimental diagnostic, especially in light of recent
work connecting X-ray-induced quenching in \thor{}:\ce{CaF2} to color-center
dynamics~\cite{Guan2025XrayQuenching}.

Because the O center yields a broad linewidth, it is not the main
clock-active quadrupole-split site considered below.
We instead focus on local compensation routes that produce a nonzero EFG and
therefore a resolved quadrupole branch pattern.
The two-F$_i$ configurations are classified as symmetry-distinct interstitial
orbits around the substitutional Th site.
The first-shell family contains 90\degree{} and 180\degree{} starting motifs,
where the labels refer to the ideal interstitial angle before relaxation; mixed
first/second-shell, second-shell, and Ca-vacancy motifs test alternative local
environments for weaker observed peaks.
Fig.~\ref{fig:motifs}(a,b) shows the two-F$_i$ structures most relevant to the
dominant and minor quadrupole-split sites; the full symmetry enumeration,
degeneracy factors, and structure gallery are given in Supplemental Sec.~I.

The calculations were performed with
VASP~\cite{Kresse1996,KresseFurthmuller1996} using the projector-augmented-wave
(PAW) method~\cite{Blochl1994,KresseJoubert1999} and the PBE
functional~\cite{Perdew1996}. A \(5\times5\times5\)
supercell of the primitive 3-atom \ce{CaF2} unit cell was used for the values
reported in Table~\ref{tab:pbe_interstitial_prl}. For the two-F$_i$ defects
this gives a 377-atom cell, \({\rm Th}_1{\rm Ca}_{124}{\rm F}_{252}\). For the
main \(90^\circ\) row, the supercell lattice vectors are fixed to the
pristine-lattice geometry. This constraint avoids folding the artificially large
periodic defect concentration into the bulk lattice vectors. The earlier
\(V_{ZZ}=223~\text{V \AA}^{-2}\) prediction used a smaller primitive
\(3\times3\times3\) supercell, corresponding to an 83-atom two-F$_i$ defect cell
\(({\rm Th}_1{\rm Ca}_{26}{\rm F}_{56})\), with relaxed lattice
vectors~\cite{Dessovic2014}; Supplemental Sec.~VI shows that these choices
strongly increase \(V_{ZZ}\) and \(\eta\). Thus the
disagreement with the earlier EFG is a convergence and lattice-constraint issue
rather than evidence against fluoride compensation: the smaller relaxed-cell
setup drives the same \(90^\circ\) motif toward the earlier high-\(V_{ZZ}\)
limit, while the larger fixed-bulk calculations recover the observed branch
scale.

EFGs were computed with spin-orbit coupling (SOC) for all tabulated structures.
The Supplemental Material documents the numerical convergence, spin-orbit
checks, sensitivity to multiple exchange-correlation
functionals~\cite{Perdew1996,Perdew2008,Armiento2005,Furness2020,Perdew2009revTPSS},
defect-energy comparisons, and extended-supercell tests that support the
production values used here.

With this computational setup, the first discriminator among local
charge-compensation geometries is their relative energy. Within the PBE
calculations, the \(90^\circ\) motif is the lowest energy member of the
same-stoichiometry two-F$_i$ family. The relative populations of these defects
may depend on growth, handling, and radiation exposure~\cite{Twidell1963}.
Nonetheless, a restricted-equilibrium estimate is obtained from $p_i(T) =
Z^{-1}g_i\exp[-\Delta E_i/(k_B T)]$, where \(g_i\) is the number of
symmetry-equivalent orientations, \(\Delta E_i\) is measured relative to the
$90^\circ$ motif, and $Z$ is the partition function.  
The resulting restricted-equilibrium populations are listed in
Table~\ref{tab:pbe_interstitial_prl} at the reported \(T=1400~{\rm K}\)
superionic-transition temperature of \ce{CaF2}.
Within this controlled comparison, the \(90^\circ\) defect is strongly favored.
The next most likely motifs are 1st-2nd obtuse, 1st-2nd acute, and
180\degree{}, all below the percent level at this temperature.

\begin{figure*}[t]
\centering
\includegraphics[width=\linewidth]{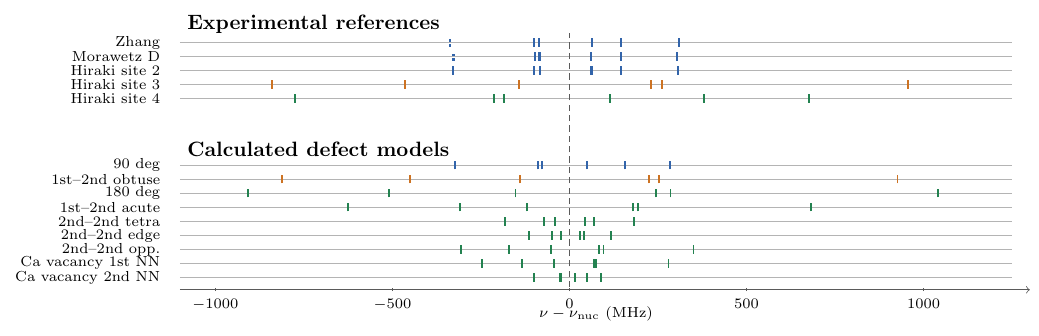}

\caption{Calculated PBE/SOC branch offsets from the \(5\times5\times5\)
primitive-cell supercell compared with experimental reference spectra. The
Zhang, Morawetz D, Hiraki site 2, and 90$^\circ$ rows are shown in blue;
the Hiraki site 3 and 1st--2nd obtuse rows are shown in orange; and the
remaining rows are shown in green. Huang \emph{et al.}
report the same zero-EFG and quadrupole-split manifold of states in independently
grown \thor{}:\ce{CaF2} crystals~\cite{Huang2026clock}. The \(90^\circ\)
two-F$_i$ motif gives the dominant quadrupole-splitting scale, whereas the
180$^\circ$ and mixed-shell motifs produce much larger splittings.}
\label{fig:branches_prl}
\end{figure*}

The second test is spectral. For each relaxed structure we diagonalize the EFG
tensor using the VASP principal-axis convention and report the two quantities
that determine the quadrupole branch pattern, \(V_{ZZ}\) and
\(\eta=(V_{YY}-V_{XX})/V_{ZZ}\), in Table~\ref{tab:pbe_interstitial_prl}. The
nuclear quadrupole interaction for a state with spin \(I\) and quadrupole moment
\(Q\) is
\begin{equation}
\hat H_Q^{(I)}
=
\frac{e Q}{2I(2I-1)}
\sum_{\alpha\beta}
V_{\alpha\beta}\hat I_\alpha \hat I_\beta,
\label{eq:hq_prl}
\end{equation}
$\alpha \in \{x,y,z\}$~\cite{CohenReif1957,Abragam1961,Slichter2013}. We
diagonalize Eq.~\eqref{eq:hq_prl} for the \(I_g=5/2\) ground state and the
\(I_e=3/2\) isomer state, using \(Q_e/Q_g=0.57003\) from the CaF$_2$ clock
measurement~\cite{Zhang2024}. The EFG splits the ground state into three Kramers
doublets and the isomer into two Kramers doublets, see Fig.~\ref{fig:motifs}(c).
The allowed and weakly mixed branches are therefore determined by \(V_{ZZ}\) and
\(\eta\), making the branch pattern a direct test of the microscopic assignment.

Fig.~\ref{fig:branches_prl} compares the experimentally measured transition
frequencies with the branch offsets predicted from the calculated EFGs. The
\(90^\circ\) two-F$_i$ motif reproduces the observed  splitting reported by Zhang \emph{et
al.}, assigned as site 2 by Hiraki \emph{et al.}, labeled the D center by
Morawetz \emph{et al.}, and used for clock operation by the PTB/Vienna group and by
Huang \emph{et
al.}~\cite{Zhang2024-Th229Comb,Hiraki2025,Morawetz2026,ToscaniDeCol2026clock,Huang2026clock}.
For the PBE/SOC \(90^\circ\) motif we find \(V_{ZZ}=107.650\) V/\AA$^2$, in good
agreement with the experimental values \(V_{ZZ}=109.1(7)\), 106.3(8), and
107.8(9) V/\AA$^2$ reported in
Refs.~\cite{Zhang2024-Th229Comb,Hiraki2025,Morawetz2026}. Calculations performed
with other functionals also match well with experiment, as shown in the
Supplemental Material.

The assignment does not rely on reproducing the asymmetry parameter exactly. The
same PBE/SOC calculation gives \(\eta=0.173\), while the experimental fits give
\(\eta=0.59163(5)\), 0.60(1), and 0.57183(9) in
Refs.~\cite{Zhang2024-Th229Comb,Hiraki2025,Morawetz2026}. The absolute branch
scale is set primarily by \(V_{ZZ}\), whereas \(\eta\) controls the detailed
nonaxial mixing and branch amplitudes. EFG tensors are local hyperfine
observables controlled by the near-nuclear charge density and are known to be
sensitive to local relaxation and electronic structure, especially in defect
environments~\cite{Petrilli1998,Errico2003,ValenzuelaReina2025,Iuliucci2023,Hartman2021}.
In the present calculations, \(\eta\) is the EFG parameter most sensitive to
supercell size, lattice-vector relaxation, and functional choice, shown in Supp.
Secs.~VI--VII. We therefore treat the static-DFT \(\eta\) mismatch as the
leading residual uncertainty, not as evidence for a thermodynamically
implausible bare dimer. The site assignment comes from the combined agreement in
overall splitting, thermodynamic preference, and restricted population.

Fig.~\ref{fig:branches_prl} also suggests that site 3 of Ref.~\cite{Hiraki2025}
is naturally associated with the 1st-2nd obtuse motif, the next most likely site in our calculations of Table~\ref{tab:pbe_interstitial_prl}. For this row the PBE/SOC
calculation gives \(V_{ZZ}=-309.049\) V/\AA$^2$ and \(\eta=0.156\), close to the
experimental site-3 values \(V_{ZZ}=-319.3(17)\) V/\AA$^2$ and \(\eta=0.15(6)\).
The reported site-3 abundance, 1--6\% of the \(90^\circ\) site, is larger than
the restricted-equilibrium estimate in Table~\ref{tab:pbe_interstitial_prl}.
This is not surprising: the table compares equilibrium
weights within a small same-stoichiometry family at the superionic-transition
scale, whereas real site fractions also depend on growth, cooling, irradiation,
and kinetic trapping. None of the local structures considered here gives a
compelling match to site 4 of Ref.~\cite{Hiraki2025}. That site may involve a
more complex defect, impurity compensation, multiple unresolved environments, or
a low-statistical-weight fit outside the present static model set.

The defect assignments are also relevant to understanding the transition
linewidth, a key parameter for nuclear clock performance. Recent measurements
show that the \ce{^229Th}:\ce{CaF2} transition linewidth depends on thorium
concentration~\cite{Ooi2026}. For a locally charge-compensated defect, a
Kanzaki-force or elastic-dipole picture is the natural long-range description:
each neutral defect complex acts as an elastic source whose strain field shifts
nearby clock nuclei, giving a concentration-dependent inhomogeneous linewidth
set by the elastic response and clock-shift sensitivity of the actual
microscopic defect~\cite{Kanzaki1957}. An uncompensated Th--Th dimer would
instead be expected to generate a more Coulomb-like long-range interaction,
potentially leading to different inhomogeneous broadening than in the neutral
defect case. Within the locally charge-compensated picture, the observed
concentration dependence is consistent with defect-induced elastic broadening,
while the zero-density intercept provides insight into the residual strain
already present in the host. As magnetic broadening is expected to be $\sim
5$~kHz~\cite{Rellergert2010}, the remaining zero-density linewidth likely
reflects residual strain in the \ce{CaF2} host, which is a well-known issue in
this material~\cite{Hirotaka2026}. This in turn motivates exploring alternative
hosts, such as \ce{ThF4}, where \ce{^229Th} incorporation does not create
compensating defects.

In summary, the dominant quadrupole-split clock site in \thor{}:\ce{CaF2} is
best assigned to a locally charge-compensated single-Th defect, not to a
dominant population of Th--Th dimers. The relaxed \(90^\circ\) two-F$_i$ motif
is thermodynamically preferred within the same-stoichiometry two-interstitial
family and gives the correct \(V_{ZZ}\) and overall splitting in the
\(5\times5\times5\) primitive-cell PBE/SOC calculation. The second most likely
configuration, the 1st-2nd obtuse motif, provides a natural assignment for a
weaker high-\(|V_{ZZ}|\) site, while the zero-EFG O center remains a separate
problem that may involve distant compensation, dynamic averaging,
radiation-induced defect conversion, or larger defect complexes. The
clock-design implication is correspondingly direct: improving the useful
quadrupole-split site fraction and linewidth should require control of local
fluoride compensation, strain, and radiation-induced defect conversion, rather
than relying on or promoting thorium aggregation. The appearance of the same
quadrupolar structure in multiple independently grown \thor{}:\ce{CaF2} crystals
reinforces this local-defect interpretation and the need to engineer the defect
chemistry of the clock-active site~\cite{Huang2026clock}.

\section*{Acknowledgments}
This work was supported by NSF awards PHYS-2013011, PHY-2207546, PHY-2412869,
PHY-2513134, and ARO award W911NF-11-1-0369. This work used Bridges-2 at
Pittsburgh Supercomputing Center through allocation PHY230110 from the ACCESS
program, which is supported by NSF grants \#2138259, \#2138286, \#2138307,
\#2137603, and \#2138296.

This work was supported by the Los Alamos National Laboratory LDRD program
project 20260021DR. Calculations were performed using resources provided by the
Los Alamos National Laboratory Institutional Computing program, including
through the Center for Integrated Nanotechnologies, a DOE-BES user facility. Los
Alamos National Laboratory, an affirmative action/equal opportunity employer, is
managed by Triad National Security, LLC, for the National Nuclear Security
Administration of the U.S. Department of Energy under contract
89233218CNA000001.

\bibliographystyle{apsrev4-2}
\bibliography{refs}

\end{document}